\documentclass[%
reprint,
superscriptaddress,
showpacs,preprintnumbers,
 amsmath,amssymb,
 aps,
prd,
]{revtex4-1}

\usepackage{graphicx}% Include figure files
\usepackage{dcolumn}% Align table columns on decimal point
\usepackage{bm}% bold math
\usepackage{bbold}
\usepackage{amssymb,amsmath}
\usepackage{hyperref}% add hypertext capabilities
\usepackage{color}
\usepackage{graphicx}
\usepackage{amsmath, amssymb}
\usepackage{amsfonts}
\usepackage{dcolumn}
\usepackage{subfigure}
\usepackage{hyperref}

\newcommand{\Tr}{\ensuremath{\operatorname{Tr}}}

\def\Fig#1{Fig.~\ref{#1}} \def\Tab#1{Tab.~\ref{#1}}
 \def\Tab#1{Tab.~\ref{#1}}

\def\Eq#1{Eq.~(\ref{#1})}
\def\eq#1{(\ref{#1})}
\def\eqref#1{(\ref{#1})}

\def\lA0{{\langle A_0 \rangle}}
\def\bA0{{\bar{A}_0}}

\def\0#1#2{\frac{#1}{#2}}

\graphicspath{{./figures/}{./}}

\begin{document}

\preprint{}

\title{Correlating the skewness and kurtosis of baryon number distributions}% Force line breaks with \\

\author{Wei-jie Fu}
% \email{W.Fu@ThPhys.Uni-Heidelberg.DE}
\affiliation{Institut f\"{u}r Theoretische Physik, Universit\"{a}t
  Heidelberg, Philosophenweg 16, 69120 Heidelberg, Germany}

\author{Jan M. Pawlowski}
% \email[]{J.Pawlowski@ThPhys.Uni-Heidelberg.DE} 
\affiliation{Institut
 f\"{u}r Theoretische Physik, Universit\"{a}t Heidelberg,
 Philosophenweg 16, 69120 Heidelberg, Germany}
\affiliation{ExtreMe Matter Institute EMMI, GSI, Planckstr. 1, 64291
  Darmstadt, Germany}


\begin{abstract}
  The skewness and the kurtosis of the baryon number distributions are
  computed within QCD-improved low energy effective models including
  quantum thermal and density fluctuations. The results are compared
  with the Beam Energy Scan experiment at RHIC. The theoretical
  results agree with the experimental measurements up to errors, for
  the collision energy $\sqrt{s}\ge 19.6\,\mathrm{GeV}$. For smaller
  collision energies a discrepancy between theoretical and
  experimental results develops. This discrepancy partially relates to
  the lack of precision of the current setup for small collision
  energies. It is outlined how this deficiency can be overcome.
\end{abstract}

                             % Classification Scheme.
\maketitle

%\tableofcontents

\section{\label{sec:intr}Introduction}

The evaluation of the QCD phase structure has
always been one of the central scientific goals in the community of
relativistic heavy ion physics \cite{Akiba:2015jwa}. Large advances
have been made in the past years, both from experimental measurements
at the Relativistic Heavy-Ion Collider (RHIC) and the Large Hadron
Collider (LHC), and theoretical calculations made in various {\it ab initio}
and effective model approaches. The QCD phase diagram is, however, far
from being unveiled. One of the most challenging tasks concerning the
phase structure of QCD is to access the existence and location of the
critical endpoint (CEP) in the phase diagram. If present, it
separates the chiral and confinement-deconfinement crossovers at low
density from the first order transition at high density, for an
overview see \cite{Stephanov:2007fk}. The Beam Energy Scan (BES)
program at RHIC is aimed at searching for the CEP of QCD, by employing
the beam energy or collision centrality dependence of the fluctuations
of conserved charges, e.g. moments of net-proton or net-charge
multiplicity distributions \cite{Adamczyk:2013dal,Luo:2015ewa}.

In response to experimental measurements, reliable theoretical
predictions of the fluctuations and correlations of conserved charges,
as well as their relation to the CEP, are highly demanded. First
principles continuum and lattice QCD calculations have made remarkable
progress in this direction in recent years, see e.g. \cite{Pawlowski:2014aha} and
\cite{Aarts:2015tyj}. At present, however,
lattice calculations are restricted to the small density regime
because of the notorious sign problem, while continuum computation have
to deal with the systematics of relevant degrees of freedom at large
densities, see e.g.\ \cite{Pawlowski:2014aha}.

Recently in \cite{Fu:2015naa}, we have investigated the QCD
thermodynamics and baryon number fluctuations within the framework of
QCD-improved low energy effective models, for related work see also
\cite{Borsanyi:2013hza,Ding:2014kva,Skokov:2010wb,Skokov:2010sf,%
  Skokov:2010uh,Karsch:2010hm,Schaefer:2011ex,Schaefer:2012gy,%
  Morita:2014fda,Morita:2014nra,Xin:2014ela,Chen:2015mga,Almasi:2016hhx}. Quantum,
thermal, and density fluctuations are embedded with the functional
renormalization group (FRG) approach to QCD, for recent developments
see e.g.
\cite{Haas:2013qwp,Herbst:2013ufa,Pawlowski:2014zaa,Helmboldt:2014iya,%
  Mitter:2014wpa,Braun:2014ata,Pawlowski:2014aha} and references
therein.

The computation of baryon number fluctuations necessitates a good
quantitative grip on the quantum, thermal, and in particular density,
fluctuations of the theory. In \cite{Fu:2015naa}, we therefore have
improved the fluctuation analysis beyond former works by including the
momentum scale dependence of the quark-meson scattering and the
nontrivial dispersions of both quarks and mesons. It was found that
the thermodynamics, e.g. the pressure and trace anomaly, and the
higher moments of baryon number distributions, after this improvement,
agree very well with the lattice results for temperatures $T\lesssim
1.2 \,T_c$. Above these temperatures the ultraviolet limitations of
the effective theory setup restrict the applicability, for a detailed
discussion see \cite{Fu:2015naa}. In this paper, we extend our
calculations to finite density, that allows a comparison to the BES
experiment at RHIC for given collision energies $\sqrt{s}$. For
smaller collision energies, $\sqrt{s}\lesssim 19.6$ GeV, that is
larger densities, similar ultraviolet limitations as for large
temperatures apply. The current effective theory setup can be
systematically improved towards QCD, e.g.\ \cite{Pawlowski:2014aha},
which then allows us to go to even smaller collision energies. This is
investigated in future work.

\section{\label{sec:effe}Effective theory setup}

\begin{figure}[t]
\centering
\includegraphics[width=8.5cm]{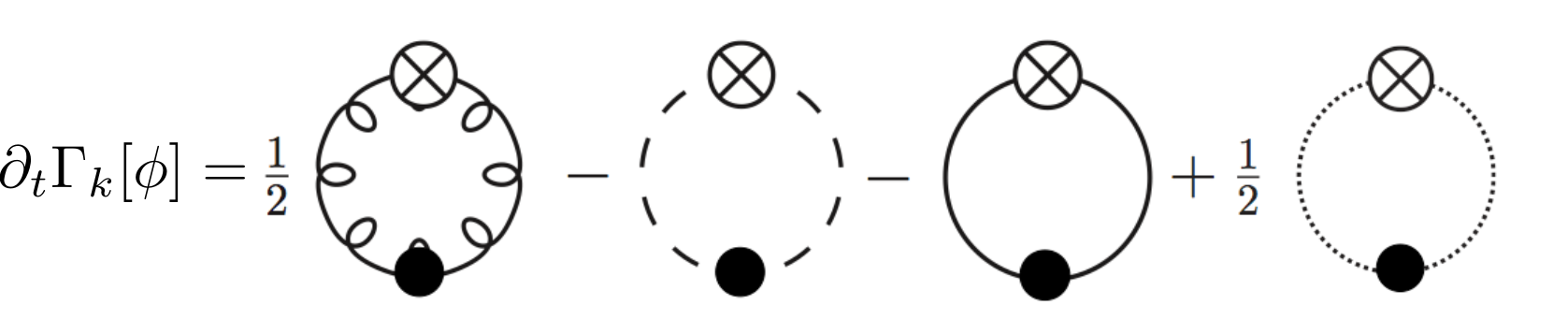}
\caption{Flow equation for the effective action $\Gamma$ in full
  QCD. The four loops correspond to contributions from gluons, ghosts,
  quarks, and mesons, respectively.}
\label{fig:FRG-QCD_rebosonised}
\end{figure}

The present effective theory is
embedded in QCD within the functional renormalization group approach.
In the present context this is described in detail in
\cite{Fu:2015naa}.  Here we give a brief summary. The flow equation
for the scale-dependent effective action of QCD, $\Gamma_k[\Phi]$,
with the super-field $\Phi=(A_\mu,c,\bar c,q,\bar q,\phi,...)$ with
$\phi=(\sigma,\vec \pi)$ and possible further effective hadronic
fields, is given by
\begin{align}
  \label{eq:dtGam}
  \partial_{t}\Gamma_{k}[\Phi]=\frac{1}{2}\mathrm{Tr}\,G_{\Phi
    \Phi}[\Phi]\partial_{t} R^{\Phi}_{k}\,, \quad t=\ln (k/\Lambda)\,,
\end{align}
with the full field-dependent propagator, 
\begin{align}\label{eq:GPhi} 
  G_{\Phi_i \Phi_j}[\Phi] =
  \left(\0{1}{\0{\delta^2\Gamma_k[\Phi]}{\delta\Phi^2}+R_k^\Phi}\right)_{ij}\,.
\end{align}
Here $k$ is the infrared cutoff scale, and $\Lambda$ is some reference
scale, typically the initial UV-scale. The flow equation \eq{eq:dtGam}
is depicted schematically in \Fig{fig:FRG-QCD_rebosonised}, for more
details and QCD results, see e.g.\
\cite{Pawlowski:2014aha,Mitter:2014wpa,Braun:2014ata}. Consequently,
the effective action can be written as
\begin{align}\label{eq:Gasplit}
  \Gamma_k[\Phi]=
  \Gamma_{\text{\tiny{glue}},k}[\Phi]+\Gamma_{\text{\tiny{matt}},k}[\Phi]\,,
  \quad
  \Gamma_{\text{\tiny{matt}},k}=\Gamma_{q,k}+
\Gamma_{\phi,k}\,,
\end{align}
where $\Gamma_{\text{\tiny{glue}},k}$ stems from the first two
diagrams in \Fig{fig:FRG-QCD_rebosonised}, and encodes the ghost- and
gluon fluctuations, while $\Gamma_{q,k}[\Phi]$ stems from the quark
loop, and $\Gamma_{\phi,k}[\Phi]$ from that of the hadronic degrees of
freedom. The split in quark and hadronic contributions, within the
framework of dynamical hadronization
\cite{Gies:2001nw,Gies:2002hq,Pawlowski:2005xe,Floerchinger:2009uf},
is a very efficient parametrization of matter
fluctuations in {\it ab initio} QCD in terms of genuine quark scatterings
and resonant momentum channels with hadronic quantum numbers, for
applications to QCD see e.g.\ \cite{Mitter:2014wpa,Braun:2014ata}.

%%%%%%%%%%%%%%%%
%%%%%%%%%%%%%%%%%%%

Integrating the QCD-flow down to scales $k=\Lambda$ with
$\Lambda\lesssim 1$ GeV, glue and ghost fluctuations start to decouple
from the matter part. This leaves us with an effective matter theory
within a glue background. These are the Polyakov-loop--extended chiral
models, such as the Polyakov--Nambu--Jona-Lasinio
model~\cite{Fukushima:2003fw,Ratti:2005jh,Fu:2007xc} and the
Polyakov--quark-meson (PQM) model~\cite{Schaefer:2007pw}. In the
current extended approximation to the PQM model setup in
\cite{Fu:2015naa}, the matter sector in \eq{eq:Gasplit} is given by
\begin{align}\nonumber 
  \Gamma_{\text{\tiny{matt}},k}&=\int_{x}\Big\{Z_{q,k}\bar{q}\big[
  \gamma_{\mu}\partial_{\mu}-\gamma_{0}(\mu+igA_0)\big]q +
  \frac{1}{2}Z_{\phi,k}(\partial_{\mu}\phi)^2\\[2ex]
  &\quad+h_{k} \,\bar{q}\left( T^{0}\sigma+i\gamma_{5}
  \vec{T}\cdot\vec{\pi}\right) q+V_{k}(\rho)-c\sigma\Big\}
  \,,
 \label{eq:action}
\end{align}
with $\int_{x}=\int_0^{1/T}d x_0 \int d^3 x$. At finite temperatures
and for constant gluonic backgrounds, only the temporal component of
the gauge field admits a nonvanishing expectation value. The meson
field $\phi=(\sigma,\vec{\pi})$ is in the $O(4)$ representation, with
$\rho=\phi^2/2$. The chemical potential $\mu$ is that of the
quark. $\vec{T}$ are the $SU(N_{f})$ generators with
$\mathrm{Tr}(T^{i}T^{j})=\frac{1}{2}\delta^{ij}$ and
$T^{0}=\frac{1}{\sqrt{2N_{f}}}\mathbb{1}_{N_{f}\times N_{f}}$. The
effective potential $V_{k}(\rho)$ is $O(4)$ invariant, and the linear
term $-c\sigma$ breaks the chiral symmetry explicitly.

%
%%%%%%%%%%%%%%%%%%%%%%%%%%%%%
\begin{figure*}[t]
\includegraphics[scale=0.35]{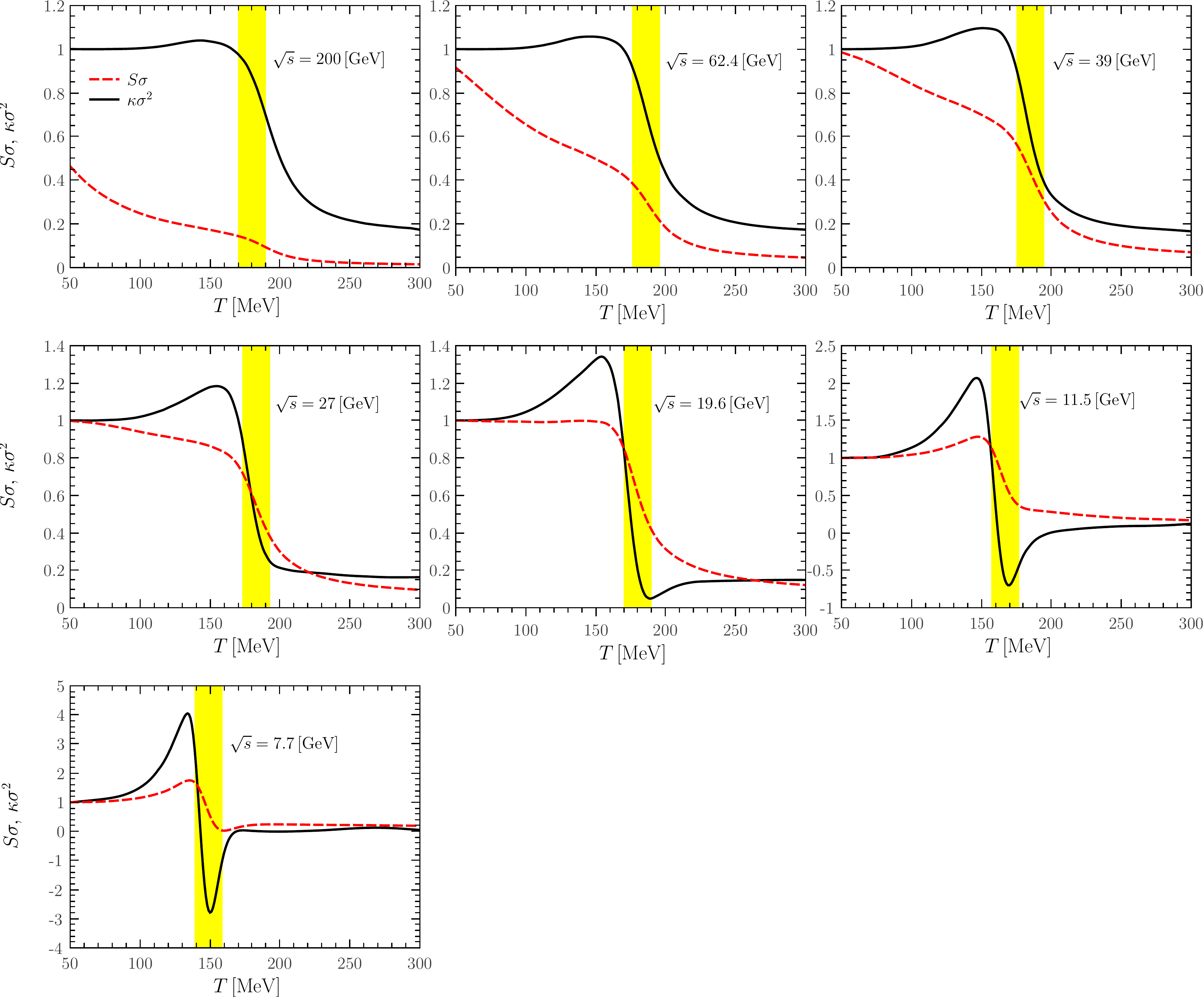}
\caption{Skewness $S\sigma$ (red dashed lines) and kurtosis
  $\kappa\sigma^2$ (black solid lines) of the baryon number
  distributions as functions of the temperature, for different values
  of chemical potential, i.e. different collision energy, shown in
  Table~\ref{tab:muB}. The yellow vertical band in each plot shows an
  estimate of the freeze-out temperature, with its center located at a
  temperature which produces the measured $S\sigma$ in
  experiments. All bands have a width of $20\,\mathrm{MeV}$ accounting
  for the systematic errors, for details see the discussion below \eq{eq:muBscale}. }\label{fig:SkewKurtT}
\end{figure*}
%%%%%%%%%%%%%%%%%%%%%%%%%
%

From \eq{eq:action} and its flow equation \eq{eq:dtGam} we extract
flow equations for the effective potential $V_{k}(\rho)$, the Yukawa
coupling $h_k$ as well as the anomalous dimensions of quark and
mesonic fields, for details see \cite{Fu:2015naa}.

As we have discussed above, for renormalization scales $k\lesssim 1$
GeV, the gluon fluctuations start to decouple from the matter
fluctuations. Hence, $\Gamma_{\text{\tiny{glue}}}$ in \eq{eq:Gasplit}
can be expressed as a functional of the gluonic background
field. Usually, it is formulated with the expectation value of the
traced Polyakov loop, viz.
\begin{align}\label{eq:Lloop}
  L(\vec{x})=\0{1}{N_c} \left\langle \Tr\, {\cal P}(\vec
    x)\right\rangle \,,\quad {\rm and }\quad  \bar L=\0{1}{N_c} \langle
  \Tr\,{\cal P}^{\dagger}(\vec x)\rangle \,,
\end{align}
with 
\begin{align}\label{eq:Ploop}
  {\cal P}(\vec x)= \mathcal{P}\exp\Big(ig\int_0^{\beta}d\tau
    A_0(\vec{x},\tau)\Big)\,.
\end{align}
Hence, $\Gamma_{\text{\tiny{glue}}}$ relates to the Polyakov loop
potential. In this work we adopt the QCD-enhanced glue potential,
\cite{Haas:2013qwp,Herbst:2013ufa}, that is also used in \cite{Fu:2015naa}. The QCD
enhancement of the potential provides the correct temperature scaling
of the glue potential in QCD. QCD-enhanced computations of
thermodynamical quantities, i.e. pressure and trace anomaly, agree
remarkably well with recent results from lattice QCD for the $N_f=2+1$
flavor case, see \cite{Herbst:2013ufa}.

Apart from the glue potential, we need to specify the initial
conditions of the flow equations for the matter sector. We choose the
UV-cutoff scale to be $\Lambda=700\,\mathrm{MeV}$ in order to maximize
the glue decoupling while still keeping as many matter fluctuations as
possible, see \cite{Fu:2015naa}. The effective potential at this
initial scale is well approximated by a classical one, which reads
\begin{align}\label{eq:VLambda}
  \bar{V}_{\Lambda}(\bar{\rho})=\frac{\bar{\lambda}_{\Lambda}}{2}\bar{\rho}^2
  +\bar{\nu}_{\Lambda}\bar{\rho}\,,
\end{align}
Additionally, we have to specify the Yukawa coupling
$\bar{h}_{\Lambda}$ and $\bar{c}_{\Lambda}$ which sets the amount of
explicit chiral symmetry breaking. These relevant couplings are fixed
by fitting hadronic observables in the vacuum, the $\pi$ decay
constant $f_\pi =92.5\,\mathrm{MeV}$, the $\pi$-meson mass
$m_{\pi}=135\,\mathrm{MeV}$, the $\sigma$-meson mass
$m_{\sigma}=450\,\mathrm{MeV}$, and the quark mass
$m_{q}=297\,\mathrm{MeV}$. The obtained values of these couplings are
$\bar{\lambda}_{\Lambda}=9.7$,
$\bar{\nu}_{\Lambda}=(0.559\,\mathrm{GeV})^2$,
$\bar{h}_{\Lambda}=7.2$, $\bar{c}_{\Lambda}=1.96\times
10^{-3}\,\mathrm{GeV}^3$. Note, that in comparison with our former
calculations \cite{Fu:2015naa} where $\bar{\nu}_{\Lambda}$ in
\eq{eq:VLambda} is chosen to be vanishing, in this work we exploit the
degree of freedom $\bar{\nu}_{\Lambda}$ to decrease meson fluctuations
above the scale of the chiral symmetry breaking
\cite{Mitter:2014wpa,Braun:2014ata}. In the same time, the updated
initial conditions reduces the mass of $\sigma$-meson to
$450\,\mathrm{MeV}$, which is consistent with $400-550\,\mathrm{MeV}$
of the mass range for $f_0(500)$ in PDG\cite{Agashe:2014kda}. We note
that a much more restricted range $446\pm 6\,\mathrm{MeV}$ of
$f_0(500)$ is also provided, for details, see \cite{Agashe:2014kda}
and reference therein.

Indeed, it has been found in \cite{Fu:2015naa} that the thermodynamics
at vanishing chemical potential, such as the pressure, trace anomaly,
baryon number fluctuations etc., obtained from FRG-calculations, agree
very well with the lattice results. With the present improved choice
of initial conditions this agreement is even better for large
temperature, as expected.

\section{\label{sec:skew}Skewness and kurtosis of the baryon number distribution}

The baryon number fluctuations are given by
\begin{equation}\label{eq:suscept}
  \chi_n^{\mathrm{B}}=\frac{\partial^n}{\partial (\mu_{\mathrm{B}}/T)^n}\frac{p}{T^4}\,,
\end{equation}
with the pressure $p=-(\Gamma_{k=0,T}[\Phi]-\Gamma_{k=0,T=0}[\Phi])$
and baryon number chemical potential $\mu_{\mathrm{B}}=3\mu$. The
$\chi_n^{\mathrm{B}}$ are related to the cumulants of baryon
multiplicity distributions: for example, the mean value $M$ is given
by $M=VT^3\chi_1^{\mathrm{B}}$, the variance is
$\sigma^2=VT^3\chi_2^{\mathrm{B}}$, the skewness is
$S=\chi_3^{\mathrm{B}}/(\chi_2^{\mathrm{B}}\sigma)$, and the kurtosis
is $\kappa=\chi_4^{\mathrm{B}}/(\chi_2^{\mathrm{B}}\sigma^2)$. Here
$V$ is the volume of the system. $S\sigma$ and $\kappa
\sigma^2$ are the volume-independent skewness and kurtosis,
respectively.

In order to compare our calculated results with experiments we need
the chemical freeze-out line, the freeze-out chemical potential and
temperature in terms of the collision energy $\sqrt{s}$. We employ the
freeze-out line in \cite{Andronic:2005yp,Andronic:2008gu}, consistent
with that in \cite{Borsanyi:2014ewa}. For more discussions about the
freeze-out conditions, see
e.g. \cite{Bazavov:2012vg,Karsch:2010ck,Alba:2014eba}. For the
experimental comparison, we have to rescale the chemical potential
and temperature in the present $N_f=2$ computations, in order to take
care of the different absolute scales. The relationship between the
absolute temperature scale in $N_f=2$ flavor and that in $N_f=2+1$,
as demonstrated in \cite{Herbst:2013ufa,Fu:2015naa}, suggests to adopt
the following linear rescaling for the chemical potential,
\begin{equation}\label{eq:muBscaleBeta}
\mu_{B,N_f=2}=\beta\,\mu_{B,N_f=2+1}\,,
\end{equation}
with the parameter $\beta\geq 1$. The absolute scale of the chemical
potential is gauged by $\mu_c$, the chemical potential of the liquid
gas transition, which is closely related to the mass scales of the
theory. This seemingly suggests $\beta=1$, as the meson and quark mass
scales are fixed to their physical value in the vacuum. Note that the
mass parameters in the model are curvature masses, and the mass
adjustment applies to the pole masses, see
\cite{Strodthoff:2011tz,Tripolt:2013jra,Helmboldt:2014iya}. It has been shown in
\cite{Helmboldt:2014iya} that the latter agree relatively well for the
mesons, but might differ for the quarks. In turn, $\beta>1$ is
suggested by the relation of the critical temperatures.

Here we determine $\beta$ as follows: for each collision energy and
given $\beta$ we employ the experimentally measured skewness
$S\sigma=\chi_3^{\mathrm{B}}/\chi_2^{\mathrm{B}}$ and the
$\sigma^2/M=\chi_2^{\mathrm{B}}/\chi_1^{\mathrm{B}}$ in
\cite{Luo:2015ewa} to determine the freeze-out temperatures, denoted
by $T_f(\chi_3^{\mathrm{B}}/\chi_2^{\mathrm{B}})$ and
$T_f(\chi_2^{\mathrm{B}}/\chi_1^{\mathrm{B}})$, respectively. For
general $\beta$ these two freeze-out temperatures differ. In the
following we fix a constant $\beta$ by the requirement
\begin{align}
T_f(\chi_3^{\mathrm{B}}/\chi_2^{\mathrm{B}})\approx 
T_f(\chi_2^{\mathrm{B}}/\chi_1^{\mathrm{B}})\,, 
\end{align}
for all collision energies. With this requirement, we obtain
$\beta=1.13$. In \Tab{tab:muB} we present the relevant
$\mu_{B,N_f=2}$, $T_f(\chi_3^{\mathrm{B}}/\chi_2^{\mathrm{B}})$ and
$T_f(\chi_2^{\mathrm{B}}/\chi_1^{\mathrm{B}})$. The slight variations
of about $5\,\mathrm{MeV}$ for the two freeze-out temperatures relate
to the constant choice of $\beta$. We emphasize, that this has little
impact on the current results. Note also that the freeze-out curve in
the present approach with its additional constraints on the freeze-out
temperature is consistent with the results in
\cite{Andronic:2005yp,Andronic:2008gu} after an appropriate rescaling.

We also note that $\beta=1.13$ is close to the $\beta$ which one gets
from simply taking over the relation of the absolute temperature
scales, to wit, 
\begin{equation}\label{eq:muBscale}
\mu_{B,N_f=2}=\frac{T_{c,N_f=2}}{T_{c,N_f=2+1}}\mu_{B,N_f=2+1}\,,
\end{equation}
where $T_{c,N_f=2}=180\,\mathrm{MeV}$ is the pseudocritical
temperature at $\mu_B=0$ for $N_f=2$, and
$T_{c,N_f=2+1}=155\,\mathrm{MeV}$ is the pseudocritical temperature
at $\mu_B=0$ for $N_f=2+1$ lattice simulations
\cite{Borsanyi:2010bp}. The latter also agrees with the freeze-out
temperature \cite{Borsanyi:2014ewa}. Substituting these values into
\eq{eq:muBscale}, one finds $\beta=1.16$. 

We proceed by discussing the systematic error in our temperature
estimate.  First of all, $
T_f(\chi_3^{\mathrm{B}}/\chi_2^{\mathrm{B}})$ and
$T_f(\chi_2^{\mathrm{B}}/\chi_1^{\mathrm{B}})$ do not have to agree
precisely, and only provide estimates for the freeze-out
temperature. The latter one also has a bigger uncertainty, but in most
regimes they differ by less than 10 MeV in an appropriate $\beta$-range. We conclude that 20 MeV accounts well for the
related systematic error leading to the yellow error bands in \Fig{fig:SkewKurtT}.

\begin{table}[t]
  \centering
  \begin{tabular}[c]{c|c|c|c|c|c|c|c}
    \hline \hline
    $\sqrt{s}\,[\mathrm{GeV}]$ & 200 & 62.4 & 39 & 27 & 19.6 & 11.5 & 7.7\\ \hline
    $\mu_{B,N_f=2}\,[\mathrm{MeV}]$&25.3&78.1&121&168.7&222.7&343&459.4\\ \hline
    $T_f(\chi_3^{\mathrm{B}}/\chi_2^{\mathrm{B}})\,[\mathrm{MeV}]$&180&186&185&183&180&167&149\\ \hline
    $T_f(\chi_2^{\mathrm{B}}/\chi_1^{\mathrm{B}})\,[\mathrm{MeV}]$&178&183&188&182&178&168&154\\ \hline
  \end{tabular}
  \caption{$\mu_{B,N_f=2}$, $T_f(\chi_3^{\mathrm{B}}/\chi_2^{\mathrm{B}})$ and  $T_f(\chi_2^{\mathrm{B}}/\chi_1^{\mathrm{B}})$ 
    corresponding to different collision energy, with $\beta=1.13$ in \Eq{eq:muBscaleBeta}. 
    Here, we have employed the data of $S\sigma$ and $\sigma^2/M$ measured 
    in $\mathrm{Au}+\mathrm{Au}$ collisions at RHIC with centrality $0-5\%$, for details see \cite{Luo:2015ewa}. }
  \label{tab:muB}
\end{table}
%

%
%%%%%%%%%%%%%%%%%%%%%%%%%%%%%
\begin{figure*}[t]
\includegraphics[scale=0.6]{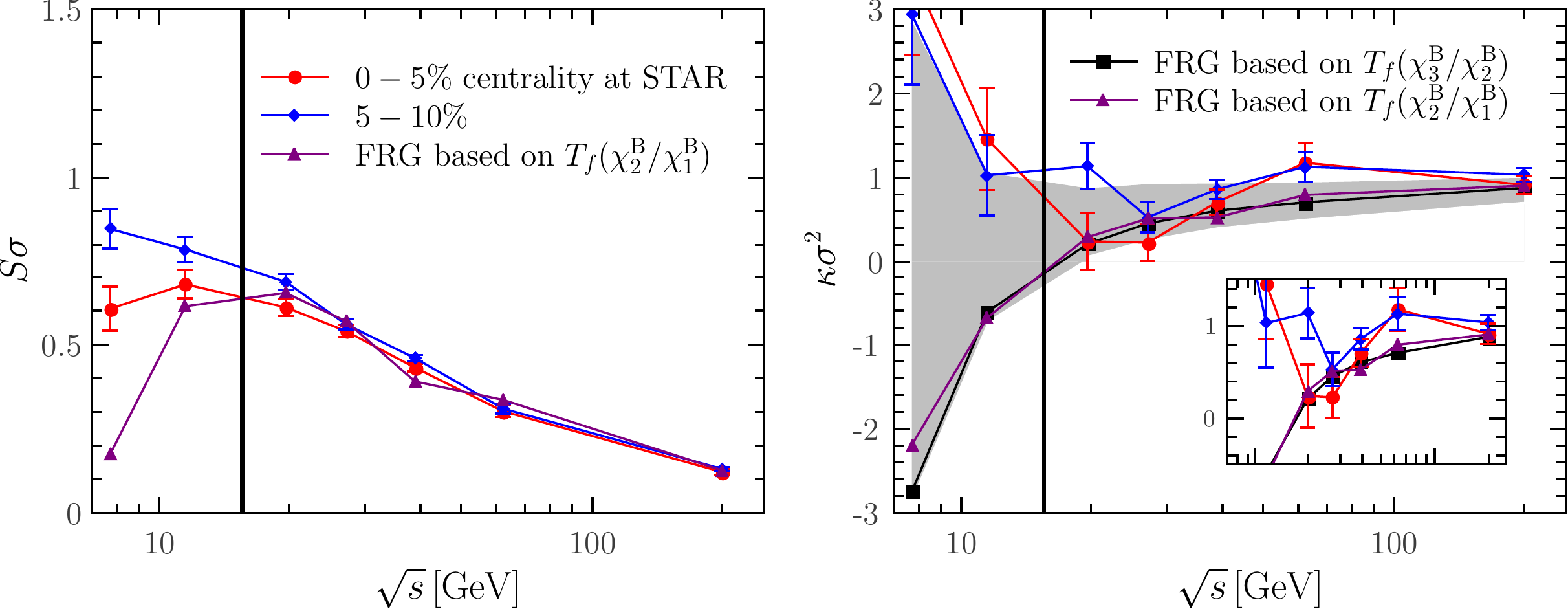}
\caption{ Theoretical calculations of $S\sigma$ (left panel) and
  $\kappa \sigma^2$ (right panel) as functions of the collision
  energy, in comparison with experimental measurements in
  $\mathrm{Au}+\mathrm{Au}$ collisions at RHIC with centralities
  $0-5\%$, $5-10\%$ \cite{Luo:2015ewa}. The gray regions show error
  estimates resulting from the determination of freeze-out
  temperatures, which corresponds to the vertical yellow bands in
  \Fig{fig:SkewKurtT}. The black vertical lines indicate the 
  collision energy, below which the UV-cutoff effect
  becomes significant.}\label{fig:SkewKurts}
\end{figure*}
%%%%%%%%%%%%%%%%%%%%%%%%%
%

\section{\label{sec:resu}Results}

To begin with, we note that the chemical potential of
electric charge $\mu_Q$ has been shown to be quite small in
comparison to $\mu_B$ \cite{Borsanyi:2013hza}, and hence is neglected
in our calculations. Furthermore, the experiments measure the moments
of net-proton multiplicity distributions. Note, that yet
  there has not been a conclusive answer about the relation between
  the net-baryon number and the net-proton number fluctuations. It was
  demonstrated in \cite{Fukushima:2014lfa} that the difference between
  them is small, if the Boltzmann approximation for the baryon
  distribution is only violated mildly. However, it is generally
  believed that in more realistic heavy-ion collisions the situation
  is more complicated , see
  e.g. \cite{Kitazawa:2011wh,Kitazawa:2012at}. Then, because of the
  global baryon number conservation, the net-baryon number and the
  net-proton number fluctuations are affected differently
  \cite{Schuster:2009jv,Bzdak:2012an}. In this work we neglect these
  intricacies for the sake of simplicity.

  In \Fig{fig:SkewKurtT} we show the skewness and the kurtosis of
  baryon number distributions as functions of temperature for
  different values of the collision energy $\sqrt{s}$. Different
  collision energy relates to different chemical potentials, see
  Table~\ref{tab:muB}. Note that the QGP produced in heavy-ion collisions evolves
  with time, and the nonequilibrium skewness and kurtosis might
  differ from equilibrium expectations, see e.g.\
  \cite{Mukherjee:2015swa}. For more discussions about the dynamical
  evolution of fluctuations, see
  e.g. \cite{Hippert:2015rwa,Jiang:2015hri,Herold:2016uvv}. Thus, it
  is expected that a range of $T$ near the freeze-out temperature,
  rather than a single value, affects experimental measurements.

  Note also that fixing the freeze-out temperature with different
  fluctuation observables does not necessarily result in identical
  results, though we have checked for $S\sigma$ and $\sigma^2/M$ that
  the differences are small, see \Tab{tab:muB}. Finally, a variation
  of the parameters (initial conditions) of the present low energy
  effective theory can be used to effectively slightly shift the
  skewness and kurtosis curves relative to each other. The ensuing
  systematic error has been discussed below \eq{eq:muBscale}, and
  leads to a loss of predictive power below 15-20 GeV collision 
  energy, signalled by the vertical line in \Fig{fig:SkewKurts}: for
  decreasing collision energy the transition strength of the
  crossover becomes larger. Then the amplitude of fluctuations,
  especially the kurtosis $\kappa\sigma^2$, increases as well, as does
  the systematic error. Nonetheless, it is remarkable that for
  collision energy less than $19.6\,\mathrm{GeV}$, the kurtosis
  $\kappa\sigma^2$ gets negative in a regime, which roughly overlaps
  with the yellow band.

  In \Fig{fig:SkewKurts} our results for $S\sigma$ and $\kappa
  \sigma^2$ are compared with experimental results by the STAR
  Collaboration at RHIC. The shaded area in \Fig{fig:SkewKurts}
  provides our systematic error estimate, which covers the variation
  ranges of $\kappa \sigma^2$ within the yellow bands in
  \Fig{fig:SkewKurtT}. Moreover, for the collision energies $\sqrt{s}<
  19.6\,\mathrm{GeV}$, the related freeze-out chemical potentials are
  bigger than $300\,\mathrm{MeV}$, and hence are of the order of the
  UV cutoff scale $\Lambda$.  Following \cite{Helmboldt:2014iya}, we
  investigate the UV limitations of the present effective theory
  setup: if the flow shows strong deviations from the vacuum flow at
  the initial scale for large $T$ and $\mu_B$, a significant part of
  the fluctuations is not encoded within the flow. Note also, that
  enlarging $\Lambda$ brings us into the regime, where glue quantum
  fluctuations become increasingly important.  Hence, a larger UV
  cutoff scale does not resolve the problem. The black vertical lines
  in \Fig{fig:SkewKurts} show the collision energy
  $\sqrt{s}\sim 15.6\,\mathrm{GeV}$, corresponding to $\mu_B\sim
  270\,\mathrm{MeV}$, below which the UV-cutoff effects become
  obvious. In other words, when the chemical potential is larger than
  $\sim 270\,\mathrm{MeV}$, quantum fluctuations of mesons above the
  scale of the UV-cutoff are not negligible any more, and thus our
  current setup has to be improved with QCD-fluctuations. This is
  deferred to future work.  

  Finally, finite volume effects \cite{Chen:2015mga}, volume
  fluctuations \cite{Skokov:2012ds} and the difference between the
  net-proton and net-baryon number fluctuations due to, e.g. the
  effect of global baryon number conservation \cite{Schuster:2009jv},
  should be also considered in a full analysis of small collision
  energies.

Recently, \cite{Luo:2015ewa}, the transverse momentum ($p_T$)
acceptance has been extended from $0.4<p_T<0.8\,\mathrm{GeV}/c$ in the
first release of experimental data in \cite{Adamczyk:2013dal} to
$0.4<p_T<2\,\mathrm{GeV}/c$. Here, we compare our calculations with
the updated $S\sigma$ and $\kappa \sigma^2$ of net-proton
distributions in $\mathrm{Au}+\mathrm{Au}$ collisions with
centralities $0\%-5\%$, $5\%-10\%$, and rapidity $|y|<0.5$
\cite{Luo:2015ewa}. These are consistent with
  recent findings that larger rapidity and $p_T$ acceptance results in a 
  significantly larger critical point signal \cite{Ling:2015yau}.

  First of all, we employ the freeze-out temperature
  $T_f(\chi_3^{\mathrm{B}}/\chi_2^{\mathrm{B}})$ in \Tab{tab:muB},
  which is determined by the experimentally measured $S\sigma$,
  i.e. the red circular points in the left panel of
  \Fig{fig:SkewKurts}. Then one can compare the computed kurtosis
  $\kappa \sigma^2$ with the experimental results, as shown by the
  black squares in the right panel. We find, that for collision
  energies $\sqrt{s}\ge 19.6\,\mathrm{GeV}$ our $\kappa \sigma^2$
  agrees with the experimental measurements within the error bars. For
  collision energies below $19.6\,\mathrm{GeV}$, the theoretical
  $\kappa \sigma^2$ drops rapidly and changes sign with the decrease
  of the energy. In turn, the experimental value of $\kappa \sigma^2$
  rises, as also shown in the inlay plot of \Fig{fig:SkewKurts}. Then,
  we use $T_f(\chi_2^{\mathrm{B}}/\chi_1^{\mathrm{B}})$ as the
  freeze-out temperature in lieu of
  $T_f(\chi_3^{\mathrm{B}}/\chi_2^{\mathrm{B}})$, and the relevant
  $S\sigma$ and $\kappa \sigma^2$ are presented in \Fig{fig:SkewKurts}
  by triangles. This provides another systematic error check of the
  present computations. We find that the results of $S\sigma$ based on
  $T_f(\chi_2^{\mathrm{B}}/\chi_1^{\mathrm{B}})$ start to deviate
  significantly from the experimental results for collision energies
  $\sqrt{s}< 19.6\,\mathrm{GeV}$. In contradistinction, $\kappa
  \sigma^2$ based on $T_f(\chi_2^{\mathrm{B}}/\chi_1^{\mathrm{B}})$ is
  consistent with that based on
  $T_f(\chi_3^{\mathrm{B}}/\chi_2^{\mathrm{B}})$ for all collision
  energies. Note however, that this simply originates in the fact that
  for low collision energy the freeze-out temperatures, both
  $T_f(\chi_3^{\mathrm{B}}/\chi_2^{\mathrm{B}})$ and
  $T_f(\chi_2^{\mathrm{B}}/\chi_1^{\mathrm{B}})$, are close to those
  corresponding to the minimum of $\kappa \sigma^2$ , as shown in
  \Fig{fig:SkewKurtT}. Consequently, theoretical curves in the right
  panel of \Fig{fig:SkewKurts} are at the bottom of the error
  estimates, and are not trustworthy anymore. 

\section{\label{sec:summ}Summary and conclusions}

  The skewness and kurtosis of baryon
  number distributions at finite temperature and chemical potential
  are calculated within a QCD-improved low energy effective
  theory. The setup includes quantum thermal and density fluctuations
  within the functional renormalization group approach. The present
  theoretical results are compared with recent experimental
  measurements. Given the skewness, our calculated kurtosis agrees
  with the experimental measurements up to errors, for the collision
  energy $\sqrt{s}\geq 19.6\,\mathrm{GeV}$. For smaller collision
  energies such a comparison is affected by several intricacies: the
  experimental data are affected by kinematic cuts, and are only
  sensitive to the net proton number. Moreover, we have not taken into
  account the nonequilibrium nature of the heavy ion
  collision. Furthermore, the present setup has to be improved
  systematically towards QCD for small collision energies, i.e.,
  pushing the initial scale of UV-cutoff to a regime of perturbative
  QCD, and including quantum fluctuations of the glue part, for more
  discussions about the full QCD in the vacuum, see
  e.g.\cite{Pawlowski:2014aha}. This extension is
  currently under progress.

\begin{acknowledgments}
  We thank X. Luo for providing us with the experimental data, and
  T.~K.~Herbst, M.~Mitter, F.~Rennecke, B.-J. Schaefer, N.~Strodthoff
  for valuable discussions and work on related subjects. W.-J. Fu
  thanks also the support of Alexander von Humboldt Foundation. This
  work is supported by EMMI, and by ERC-AdG-290623.
\end{acknowledgments}

%\nocite{*}

%\bibliography{refspec}% Produces the bibliography via BibTeX.
\bibliography{ms.bib}% Produces the bibliography via BibTeX.

\end{document}